\documentclass[12pt]{article}

\usepackage{latexsym}
\usepackage{amssymb,amsfonts,amsmath}
\usepackage{graphicx} 
\usepackage{indentfirst}
\usepackage{bbm}
\usepackage{amssymb}
\usepackage{verbatim}
\usepackage{amsmath, amsthm,amssymb}
\usepackage{mathrsfs}
\usepackage{hyperref}
\usepackage{amsfonts}
\usepackage{dsfont}
\usepackage{cite}
\usepackage{xcolor}
\usepackage[multiple]{footmisc}

\parskip=\medskipamount

\newcommand {\cD}{{\cal D}}
\newcommand {\cE}{{\cal E}}

\newcommand {\cH}{{\cal H}}

\newcommand {\cJ}{{\cal J}}
\newcommand {\cK}{{\cal K}}
\newcommand {\cL}{{\cal L}}
\newcommand {\cM}{{\cal M}}
\newcommand {\cN}{{\cal N}}
\newcommand {\cO}{{\cal O}}

\newcommand {\cS}{{\cal S}}

\newcommand {\cU}{{\cal U}}
\newcommand {\cV}{{\cal V}}

\def\a{\alpha}

\def\b{\beta}

\def\d{\delta}

\def\g{\gamma}
\def\G{\Gamma}

\def\l{\lambda}
\def\m{\mu}

\def\p{\pi}
\def\q{\theta}

\def\x{\xi}
\def\z{\zeta}
\def\D{\Delta}
\def\F{\Phi}
\def\J{\Psi}
\def\L{\Lambda}
\def\O{\Omega}

\def\S{\Sigma}
\def\U{\Upsilon}

\def\rd{{\rm d}}
\def\ri{{\rm i}}
\def\re{{\rm e}}

\newcommand{\ad}{{\dot{\alpha}}}                           %
\newcommand{\bd}{{\dot{\beta}}}                            %
\newcommand{\ve}{\varepsilon}                            %

\renewcommand{\aa}{{\a\ad}}
\newcommand{\bb}{{\b\bd}}
\newcommand{\pa}{\partial}                           %
\newcommand{\hf}{\frac12}

\newcommand{\be}{\begin{equation}}
\newcommand{\ee}{\end{equation}}
\newcommand{\bea}{\begin{eqnarray}}
\newcommand{\eea}{\end{eqnarray}}
\newcommand{\non}{\nonumber}
\newcommand{\1}{{\underline{1}}}
\newcommand{\2}{{\underline{2}}}

\newcommand{\bm}[1]{\mbox{\boldmath$#1$}}

\def\double #1{#1{\hbox{\kern-2pt $#1$}}}

\newcommand{\gd}{{\dot\g}}
\newcommand{\dd}{{\dot\d}}

\newcommand{\dalpha}{{\dot{\alpha}}}

\newcommand{\N}{{\mathcal N}}

\newif\ifdtup

\def\de{{\nabla}}                                         %

\newcommand{\bsubeq}{\begin{subequations}}
\newcommand{\esubeq}{\end{subequations}}

\newcommand{\eol}{\notag \\}

\numberwithin{equation}{section}

\newcommand{\sSU}{\mathsf{SU}}

\newcommand{\sU}{\mathsf{U}}

\makeatletter
\def\widebreve{\mathpalette\wide@breve}
\def\wide@breve#1#2{\sbox\z@{$#1#2$}%
	\mathop{\vbox{\m@th\ialign{##\crcr
				\kern0.08em\brevefill#1{0.8\wd\z@}\crcr\noalign{\nointerlineskip}%
				$\hss#1#2\hss$\crcr}}}\limits}
\def\brevefill#1#2{$\m@th\sbox\tw@{$#1($}%
	\hss\resizebox{#2}{\wd\tw@}{\rotatebox[origin=c]{90}{\upshape(}}\hss$}
\makeatletter

\begin{document}

\begin{titlepage}
\begin{flushright}
July, 2024 \\
Revised version: October, 2024
\end{flushright}
\vspace{5mm}

\begin{center}
{\Large \bf Towards $\mathcal{N}=2$ superconformal higher-spin theory}
\end{center}

\begin{center}

{\bf Sergei M. Kuzenko and Emmanouil S. N. Raptakis} \\
\vspace{5mm}

\footnotesize{
{\it Department of Physics M013, The University of Western Australia\\
35 Stirling Highway, Perth W.A. 6009, Australia}}  
~\\
\vspace{2mm}
~\\
Email: \texttt{ 
sergei.kuzenko@uwa.edu.au, emmanouil.raptakis@uwa.edu.au}\\
\vspace{2mm}

\end{center}

\begin{abstract}
\baselineskip=14pt
Three years ago, we proposed free off-shell models for ${\mathcal N}=2$ superconformal higher-spin multiplets in arbitrary conformally flat backgrounds, constructed conserved conformal higher-spin supercurrents for a massless hypermultiplet, and sketched the Noether procedure to generate its cubic couplings to the superconformal higher-spin multiplets. 
This paper is devoted to completing the Noether procedure. Specifically, we: (i) describe the unique off-shell primary extensions of the conformal higher-spin supercurrents; (ii) embed the off-shell superconformal prepotentials 
into primary unconstrained isotwistor multiplets; and (iii) present the unique gauge transformations of the hypermultiplet and the isotwistor prepotentials. An extension of the Noether procedure beyond the cubic level is also sketched, following the earlier ${\mathcal N}=1$ superconformal approach 
developed by the authors and Ponds in 2022.
Our construction is based on making use of the polar hypermultiplet within the projective-superspace setting.
\end{abstract}
\vspace{5mm}

\vfill

\vfill
\end{titlepage}

\newpage
\renewcommand{\thefootnote}{\arabic{footnote}}
\setcounter{footnote}{0}

\tableofcontents{}
\vspace{1cm}
\bigskip\hrule

\allowdisplaybreaks

\section{Introduction}

Three years ago, our work \cite{KR21}  presented free off-shell models for $\cN=2$ superconformal higher-spin multiplets in arbitrary conformally flat backgrounds. We also described all the  $\N=2$ conformal supercurrent multiplets $J^{\a(m) \ad(n)}$, with $m$ and $n$ non-negative integers.\footnote{It should be noted that not all $\cN=2$ superconformal multiplets of conserved currents are of this form. In particular, the flavour current multiplet 
is described by a real iso-triplet $J^{ij}$, the linear multiplet \cite{BS,SSW}, while the source for the $\cN=2$ superconformal gravitino multiplet is an isospinor $J^i$ \cite{HKR}.}
The $m=n=0$ case corresponds to the ordinary conformal supercurrent $J$ 
\cite{Sohnius, HST}, which is the source of the Weyl multiplet of $\cN=2$ conformal supergravity
\cite{BdeRdeW}. The supergravity origin of $J$ and its non-conformal extensions was uncovered in \cite{KT,BK2,Butter:2011ym}. 

For a massless hypermultiplet, the conserved conformal supercurrents
$J^{\a(s) \ad(s)}$, with $s\geq 0$, were derived in \cite{KR21} for an arbitrary conformally flat background. In the $s=0$ case, the conserved  supercurrent $J$  exists in an arbitrary supergravity background
 \cite{KT}. It should be stressed that each conserved conformal hypermultiplet supercurrent $J^{\a(s) \ad(s)}$ is uniquely defined by the corresponding conservation equation. This equation implies, in particular, that $J^{\a(s) \ad(s)}$ is a conformal primary superfield.
 
 In \cite{KR21} we sketched the Noether procedure to generate cubic couplings of the hypermultiplet to the superconformal higher-spin multiplets. The present paper is aimed at completing this procedure. First of all, we provide off-shell extensions of the conserved hypermultiplet supercurrents $J^{\a(s) \ad(s)}$ given in \cite{KR21}. For this we make use of the curved projective-superspace formalism of 
 \cite{KLRT-M1, KLRT-M2} 
 (which is a natural extension of the earlier five-dimensional formulation (5D) of 
 \cite{Kuzenko:2007hu, Kuzenko:2008wr}) 
 in conjunction with the conformal superspace methods for supersymmetric theories with eight supercharges in four and five dimensions \cite{ButterN=2, BKNT-M15}.\footnote{The conformal superspace approach to supergravity-matter systems was pioneered by Butter in four dimensions for the $\cN=1$ \cite{ButterN=1} and $\cN=2$ \cite{ButterN=2} cases. Subsequently it has been extended to two \cite{Kuzenko:2022qnb},  three \cite{BKNT-M1, BKNT-M2, KNT-M},  five \cite{ BKNT-M15} and six \cite{BKNT} dimensions. One may argue that the problem of deriving the complete actions for $\cN=4$ conformal supergravity in four dimensions (involving a holomorphic function of the complex scalar that parametrises an $\sSU(1, 1) / \sU(1)$ coset space), which  were constructed only a few years ago \cite{Butter:2016mtk, Butter:2019edc}, was solved using the $\cN=4$ conformal superspace
sketched in the appendices of \cite{Butter:2019edc}. The $\cN=3$ conformal superspace in four dimensions has recently been developed \cite{Kuzenko:2023qkg}.}  
 In particular, our definitions of covariant projective and isotwistor multiplets follow \cite{BKNT-M15}
 (in the 4D  case, these definitions are reviewed in \cite{KRTM}).\footnote{It follows from the analysis in \cite{KRTM} that these definitions are equivalent to those given \cite{KLRT-M1, KLRT-M2}.}
 Secondly, we embed the off-shell superconformal prepotentials of \cite{KR21} into primary unconstrained isotwistor multiplets. Thirdly, we present unique gauge transformations for the hypermultiplet and the isotwistor prepotentials.

This work opens up a new direction within the framework of conformal higher-spin (CHS) theory. The first influential work on CHS theory was written in 1985 by Fradkin and Tseytlin \cite{FT} who proposed free gauge-invariant actions  for CHS gauge fields (spin $s>2$) in Minkowski space ${\mathbb M}^4$ as a generalisation of the models for massless spin-1 gauge field, conformal gravitino (spin 3/2) and conformal graviton (spin 2). Cubic couplings for CHS fields of all integer spins $s\geq 2$ were constructed by Fradkin and Linetsky in 1989 \cite{FL}, and a year later  their results were extended 
 to the superconformal case \cite{FL-4D}. The Fradkin-Linetsky approach was based on gauging the infinite-dimensional CHS algebra  $\mathfrak{hsc}^\infty (4) $ introduced in \cite{FL-algebras}  and its superconformal extension  $\mathfrak{shsc}^\infty (4|1) $ also constructed in \cite{FL-algebras}.  Such superalgebras provide a natural extension of 
 the  anti-de Sitter (AdS) higher-spin superalgebras pioneered by Fradkin and Vasiliev \cite{FV1,FV2,Vasiliev88}.

So far the Fradkin-Linetsky geometric formalism in four dimensions \cite{FL-4D} has not been extended beyond the cubic approximation. An alternative approach was put forward by Segal in 2002 \cite{Segal} (soon after the powerful
proposal by Tseytlin \cite{Tseytlin}) who constructed the unique bosonic gauge theory of interacting symmetric traceless tensor fields of all ranks in a spacetime of any even dimension $d\geq 4$. Nowadays, Segal's theory is often referred to as CHS gravity, see e.g. \cite{BBJ,Bekaert:2022poo, Basile:2022nou}. As discussed in \cite{Basile:2022nou}, 
there are two constructions of CHS gravity: (i) the induced action approach
advocated by Tseytlin \cite{Tseytlin} and elaborated on in \cite{BJM}; and (ii) Segal's approach \cite{Segal}, which deals with a particle  model and deformation quantisation. Although the two constructions are equivalent (see the discussion in \cite{Basile:2022nou}), it is the induced action approach which can be naturally generalised to the supersymmetric case. 

In their gauging of the CHS superalgebra $\mathfrak{shsc}^\infty (4|1) $, Fradkin and Linetsky \cite{FL} did not identify off-shell $\cN=1$ superconformal analogues of the free CHS models \cite{FT}. Free $\cN=1$ superconformal higher-spin (SCHS) gauge theories were constructed in 2017 in the Minkowski and anti-de Sitter superspaces \cite{KMT}, and later also in arbitrary conformally-flat backgrounds \cite{KP, Kuzenko:2020jie}.
Recently, Ref. \cite{KPR}  derived, for the first time, the model for a conformal scalar/chiral multiplet  coupled to an infinite set of background higher-spin superfields, thus providing the complete setup for developing an induced action approach to  $\cN=1$ SCHS gravity.  This approach has been utilised to compute the leading-order contribution to the nonlinear $\cN=1$ SCHS action in \cite{KLaP}. The present paper is aimed at extending the results of 
\cite{KPR} to the $\cN=2$ superconformal case. 

There exist two fully-fledged superspace approaches to formulate off-shell $\cN=2$ supersymmetric field theories: (i) harmonic superspace \cite{GIKOS,GIOS};  and (ii) projective superspace \cite{KLR,LR1,LR2}.
In the rigid supersymmetric case, they make use of the same superspace 
\bea
{\mathbb M}^{4|8}\times {\mathbb C}P^1={\mathbb M}^{4|8}\times S^2
\eea
which was introduced for the first time by Rosly \cite{Rosly}.
However, they differ in the following: (i) the structure of off-shell   supermultiplets used; 
and (ii) the supersymmetric action principle chosen.\footnote{The relationship between the {\it rigid} harmonic and projective superspace formulations is spelled out in \cite{K_double, Jain:2009aj, Kuzenko:2010bd, Butter:2012ta}.} 
In particular, they deal with different off-shell realisations for the so-called charged hypermultiplet:
(i) the $q^+$ hypermultiplet \cite{GIKOS} in harmonic superspace; and (ii) the polar hypermultiplet \cite{LR1} in projective superspace.\footnote{The terminology ``polar hypermultiplet'' was introduced in the influential paper \cite{G-RRWLvU}.}

Our construction in this paper is based on making use of the covariant polar hypermultiplet
 \cite{KLRT-M1, KLRT-M2, Kuzenko:2007hu, Kuzenko:2008wr}
 in a conformally flat  projective superspace. Recently, an alternative harmonic-superspace construction was advocated in \cite{Buchbinder:2024pjm} to formulate cubic couplings of the $q^+$ hypermultiplet to $\mathcal{N} = 2$ superconformal higher-spin multiplets in Minkowski superspace.\footnote{The construction of \cite{Buchbinder:2024pjm} was based on the earlier work by the same authors \cite{Buchbinder:2021ite, Buchbinder:2022kzl, Buchbinder:2022vra} concerning the description of massless harmonic higher-spin multiplets and their cubic couplings to the $q^+$ hypermultiplt.}  We will provide comments on the approach of \cite{Buchbinder:2024pjm} in section \ref{section4.3}.

This paper is organised as follows. In order to make this work reasonably self-contained, the next two sections include review material. 
Specifically, section \ref{AppendixCSS} provides the salient details of $\cN=2$ conformal superspace pertinent to this work, while  
section \ref{AppendixA} reviews those concepts of curved projective superspace which are used in later sections. 
In section \ref{IsotwistorCSCs}
we present unique 
off-shell extensions of the conserved conformal supercurrents proposed in \cite{KR21}. 
A superfield Noether procedure is employed in section \ref{Section3} 
to engineer manifestly superconformal interactions between a polar hypermultiplet and an infinite tower of superconformal higher-spin multiplets. 
A summary of our results, as well as a sketch of possible extensions of this work, 
is provided in section \ref{Section4}. The main body of this paper is accompanied by a technical appendix.  %
Here, we collect the main results of \cite{KR21}, which are expanded upon in the main body.

Throughout this paper we employ the spinor conventions of \cite{Book}, which are similar to those of \cite{WB}, and make use of the convention that indices denoted by the same symbol are to be symmetrised over, e.g. 
\begin{align}
	U_{\a(m)} V_{\a(n)} = U_{(\a_1 . . .\a_m} V_{\a_{m+1} . . . \a_{m+n})} =\frac{1}{(m+n)!}\big(U_{\a_1 . . .\a_m} V_{\a_{m+1} . . . \a_{m+n}}+\cdots\big)~. \label{convention}
\end{align}
In sections \ref{AppendixCSS} and  \ref{AppendixA}, we deal with $\cN=2$ conformal supergravity and its matter couplings. For the remainder of the paper,  
we work solely in conformally flat backgrounds, which are characterised by vanishing super-Weyl tensor \eqref{superWeyl}
\begin{align}
	W_{\a \b} = 0~.
\end{align}

\section{$\mathcal{N}=2$ conformal superspace in four dimensions}\label{AppendixCSS}

In this section, we give the salient details of $\mathcal{N} = 2$ conformal superspace, a formulation for off-shell $\mathcal{N}=2$ conformal supergravity developed by Butter \cite{ButterN=2}, reformulated in \cite{BN}, and recently reviewed in \cite{KRTM}.

We consider a curved $\cN=2$ superspace $\mathcal{M}^{4|8}$
parametrised by local coordinates 
$z^{M} = (x^{m},\theta^{\m}_\imath,\bar \theta_{\dot{\mu}}^\imath)$, where $m = 0,1,2,3$, $\m = 1,2$, $\dot{\m} = \dot{1}, \dot{2}$, and $\imath = \underline{1}, \underline{2}$. 
Its structure group is chosen to be the $\cN=2$ superconformal group, $\sSU(2,2|2)$. The corresponding superalgebra is spanned by the supertranslation $P_A=(P_a, Q_\a^i ,\bar Q^\ad_i)$, Lorentz $M_{ab}$, dilatation $\mathbb{D}$,  R-symmetry $\mathbb{Y}$ and $J^{ij}$, and the special superconformal $K^A=(K^a, S^\a_i ,\bar S_\ad^i)$ generators. The covariant derivatives $\nabla_A = (\nabla_a, \nabla_\alpha^i, \bar\nabla^\dalpha_i)$ then take the form
\begin{align}
	\nabla_A &= E_A - \hf \Omega_A{}^{bc} M_{bc} - \Phi_A{}^{jk} J_{jk} - \ri \Phi_A \mathbb{Y}
	- B_A \mathbb{D} - \frak{F}_{A B} K^B \eol
	&= E_A - \Omega_A{}^{\b\g} M_{\b\g} - \bar{\Omega}_A{}^{\bd\gd} \bar{M}_{\bd\gd}
	- \Phi_A{}^{jk} J_{jk} - \ri \Phi_A \mathbb{Y} - B_A \mathbb{D} - \frak{F}_{A B} K^B ~.
	\label{A.1}
\end{align}
Here $E_A = E_A{}^M \pa_M$ is the supervielbein, $\Omega_A{}^{bc}$ the Lorentz connection,
and $\Phi_A{}^{jk}$ and $\Phi_A$ are the $\sSU(2)_R$ and $\sU(1)_R$ connections,
respectively. In addition, we have a dilatation connection $B_A$ and a special
superconformal connection $\frak F_{AB}$.

The Lorentz ($M_{ab}$) and $\sSU(2)_R$ ($J^{ij}$) generators are defined to act on Weyl spinors, vectors and isospinors in the following way:
\begin{subequations}
	\begin{align}
		M_{\a \b} \psi_\g &= \ve_{\g (\a} \psi_{\b)} ~, \quad \bar M_{\ad \bd} \bar \psi_\gd = \ve_{\gd (\ad} \psi_{\bd)} ~, \\
		M_{ab} V_c &= 2 \eta_{c [a} V_{b]} ~, \quad J^{ij} \chi^k = \ve^{k(i} \chi^{j)} ~.
	\end{align}
\end{subequations}
The $\sU(1)_R$ and dilatation generators obey:
\begin{subequations}
	\label{N=2ConfAlg}
	\begin{align}
		[\mathbb{Y}, \nabla_\a^i] &= \nabla_\a^i ~,\quad [\mathbb{Y}, \bar\nabla^\ad_i] = - \bar\nabla^\ad_i~,  \non \\
		[\mathbb{D}, \nabla_a] &= \nabla_a ~, \quad
		[\mathbb{D}, \nabla_\a^i] = \hf \nabla_\a^i ~, \quad
		[\mathbb{D}, \bar\nabla^\ad_i] = \hf \bar\nabla^\ad_i ~.
	\end{align}
	On the other hand, the special superconformal generators $K^A = (K^a, S^\alpha_i, \bar S_\dalpha^i)$ carry opposite $\sU(1)_R$ charge and dilatation weight to $\nabla_A$:
	\begin{align}
		[\mathbb{Y}, S^\a_i] &= - S^\a_i ~, \quad
		[\mathbb{Y}, \bar{S}^i_\ad] = \bar{S}^i_\ad~, \non \\
		[\mathbb{D} , K_a] &= - K_a ~, \quad
		[\mathbb{D}, S^\a_i] = - \hf S^\a_i ~, \quad
		[\mathbb{D}, \bar{S}_\ad^i] = - \hf \bar{S}_\ad^i ~.
	\end{align}
	Among themselves, the special superconformal generators $K^A$ obey the algebra
	\begin{align}
		\{ S^\a_i , \bar{S}^j_\ad \} &= 2 \ri \d^j_i K^\a{}_{\ad}~,
	\end{align}
	with all the other (anti-)commutators vanishing. Finally, the algebra of $K^A$ with $\nabla_B$ is
	\begin{align}
		[K_\aa, \nabla_\bb] &= -4 \ve_{\a\b} \ve_{\ad \bd} \mathbb{D} + 4 \ve_{\ad \bd} M_{\a \b} + 4 \ve_{\a \b} \bar M _{\ad \bd} ~,\non \\
		\{ S^\a_i , \nabla_\b^j \} &= \d^j_i \d^\a_\b (2 \mathbb{D} - \mathbb{Y}) - 4 \d^j_i M^\a{}_\b + 4 \d^\a_\b J_i{}^j ~,\non \\
		\{ \bar{S}^i_\ad , \bar{\nabla}^\bd_j \} &= \d^i_j \d^\bd_\ad (2 \mathbb{D} + \mathbb{Y})
		+ 4 \d^i_j \bar{M}_\ad{}^\bd - 4 \d_\ad^\bd J^i{}_j ~,\non \\
		[K_\aa, \nabla_\b^j] &= - 2 \ri \ve_{\a \b} \bar{S}_\ad^j \ , \quad [K_\aa, \bar{\nabla}^\bd_j] = 
		- 2\ri \d_\ad^\bd S_{\a j} ~, \non \\
		[S^\a_i , \nabla_\bb] &= 2 \ri \d^\a_\b \bar{\nabla}_\bd^i \ , \quad [\bar{S}^i_\ad , \nabla_\bb ] = 
		- 2 \ri \ve_{\ad \bd} \nabla_{\b i} \ ,
	\end{align}
\end{subequations}
where all other graded commutators vanish.

By definition, the gauge group of conformal supergravity  is generated by local transformations of the form
\begin{subequations}\label{SUGRAtransmations}
	\bea
	\delta_\cK \nabla_A &=& [\cK,\nabla_A] ~, \\
	\cK &=&  \xi^B \nabla_B+ \hf \L^{bc} M_{bc} + \S \mathbb{D} + \ri \rho \mathbb{Y}
	+ \L^{jk} J_{jk}
	+ \L_B K^B ~,
	\eea
\end{subequations}
where  the gauge parameters satisfy natural reality conditions.
The supergravity gauge group acts on a conformal tensor superfield $\J$ (with indices suppressed) as 
\bea 
\d_{\cK} \J = \cK \J ~.
\eea
We say that $\J$ is primary if it is annihilated by the special superconformal generators
\bea
K^A \Psi = 0 ~.
\eea
Additionally, its dimension $\D_\Psi$ and $\sU(1)_R$ charge $q_\Psi$ are defined as follows:
\bea
\mathbb{D} \Psi = \D_\Psi \Psi ~, \quad \mathbb{Y} \Psi = q_\Psi \Psi ~.
\eea
Of particular importance are primary covariantly chiral superfields, which satisfy
\bea
K^A \J=0~, \qquad \bar{\nabla}_{\ad}^i \J = 0~.
\eea
The consistency of these constraints with the superconformal algebra described above leads to highly non-trivial implications. In particular, $\J$ can carry no isospinor or dotted spinor indices, $\Psi = \Psi_{\a(m)}$, and its $\sU(1)_R$ charge and dimension are related as follows:
\bea
\label{chiralDimCharge}
q_\Psi = - 2 \D_\Psi ~.
\eea
Further, we note that for any primary tensor superfield $\Phi_{\a(m)}$ with the property $q_\Phi = - 2 \D_\Phi$, the following object 
\bea
\Psi_{\a(m)} = \bar{\nabla}^4 \Phi_{\a(m)} := \frac{1}{48} \bar{\nabla}^{ij} \bar{\nabla}_{ij} \Phi_{\a(m)}
\eea
is both chiral and primary \cite{KTM08}. Here we have made the definitions 
\begin{align}
	\qquad {\nabla}^{ij} := {\nabla}^{\a(i} {\nabla}_\a^{ j)}~, \qquad \bar{\nabla}^{ij} = \bar{\nabla}_{\ad}^{(i} \bar{\nabla}^{\ad j)}~.
\end{align}

In \cite{ButterN=2} it was shown that, in order to reproduce the component structure of conformal supergravity, certain constraints must be imposed on the graded commutators $[\nabla_A , \nabla_B \}$. In particular, they must be expressed solely in terms of the super-Weyl tensor, $W_{\a(2)}$,
\bea
\label{superWeyl}
K^B W_{\a \b}  = 0 ~, \quad \bar{\nabla}_\ad W_{\a \b} = 0 ~, \quad \mathbb{D} W_{\a \b} = W_{\a \b}~,
\eea
and its covariant derivatives. The solution to the aforementioned constraints is given by
\begin{subequations}\label{CSGAlgebra}
	\begin{align}
		\{ \nabla_\a^i , \nabla_\b^j \} &= 2 \ve^{ij} \ve_{\a\b} \bar{W}_{\gd\dd} \bar{M}^{\gd\dd} + \hf \ve^{ij} \ve_{\a\b} \bar{\nabla}_{\gd k} \bar{W}^{\gd\dd} \bar{S}^k_\dd - \hf \ve^{ij} \ve_{\a\b} \nabla_{\g\dd} \bar{W}^\dd{}_\gd K^{\g \gd}~, \\
		\{ \nabla_\a^i , \bar{\nabla}^\bd_j \} &= - 2 \ri \d_j^i \nabla_\a{}^\bd~, \\
		[\nabla_{\a\ad} , \nabla_\b^i ] &= - \ri \ve_{\a\b} \bar{W}_{\ad\bd} \bar{\nabla}^{\bd i} - \frac{\ri}{2} \ve_{\a\b} \bar{\nabla}^{\bd i} \bar{W}_{\ad\bd} \mathbb{D} - \frac{\ri}{4} \ve_{\a\b} \bar{\nabla}^{\bd i} \bar{W}_{\ad\bd} Y + \ri \ve_{\a\b} \bar{\nabla}^\bd_j \bar{W}_{\ad\bd} J^{ij}
		\eol & \quad
		- \ri \ve_{\a\b} \bar{\nabla}_\bd^i \bar{W}_{\gd\ad} \bar{M}^{\bd \gd} - \frac{\ri}{4} \ve_{\a\b} \bar{\nabla}_\ad^i \bar{\nabla}^\bd_k \bar{W}_{\bd\gd} \bar{S}^{\gd k} + \frac{1}{2} \ve_{\a\b} \nabla^{\g \bd} \bar{W}_{\ad\bd} S^i_\g
		\eol & \quad
		+ \frac{\ri}{4} \ve_{\a\b} \bar{\nabla}_\ad^i \nabla^\g{}_\gd \bar{W}^{\gd \bd} K_{\g \bd}~.
	\end{align}
\end{subequations}	
We also find that $W_{\a \b}$ must satisfy the Bianchi identity
\begin{align}
	B = \nabla_{\a\b} W^{\a\b} &= \bar{\nabla}^{\ad\bd} \bar{W}_{\ad\bd}  = \bar{B} ~,
\end{align}
where $B$ is the $\N=2$ super-Bach tensor.

To conclude this section, we point out that the equation of motion for $\cN=2$ conformal supergravity is the super-Bach-flatness condition
\bea
B = 0~.
\eea

\section{Rudiments of curved projective superspace}
\label{AppendixA} 

The conformal superspace geometry described above 
is a powerful framework for the study of superconformal field theories, but it is insufficient for the consideration of off-shell charged hypermultiplets. To circumvent this, we extend our supermanifold to
\begin{align}
	\cM^{4|8} \rightarrow \cM^{4|8} \times \mathbb{C}P^1~,
\end{align}
yielding curved projective superspace. In practice, the coordinates of $\cM^{4|8}$ are appended by an isotwistor $v^i \in \mathbb{C}^2 \setminus \{0\}$ defined modulo the equivalence relation $v^i \sim \mathfrak{c} v^i$, ${\mathfrak c} \in \mathbb{C}\setminus  \{0\}$. It is also useful to introduce a second isotwistor $u^i \in \mathbb{C}^2 \setminus \{0\}$ which is linearly independent of $v^i$, $(v,u) := v^i u_i \neq 0$, but otherwise arbitrary.

\subsection{Superconformal projective multiplets}

In accordance with  \cite{KLRT-M2,KRTM}, 
a {\it superconformal projective multiplet}\footnote{The concept of superconformal projective multiplets was originally introduced within the framework of rigid supersymmetry in \cite{K2006} and then extended to conformal supergravity in five dimensions \cite{Kuzenko:2008wr}.}
of weight $n$,
$Q^{(n)}(z,v)$, is a primary superfield on $\cM^{4|8}$ with respect to the superspace coordinates $z^A$, is holomorphic with respect to the isotwistor variables $v^i$ on an open domain of 
${\mathbb C}^2 \setminus  \{0\}$, 
and is characterised by the following conditions:

(i) it obeys the analyticity constraints 
\begin{subequations}
	\bea
	\nabla^{(1)}_{\a} Q^{(n)} := v_i \nabla_\a^i Q^{(n)} =0~, \qquad {\bar \nabla}^{(1)}_{\ad} Q^{(n)} :=  v_i \bar{\nabla}_\ad^i Q^{(n)} =0~;
	\label{ana}
	\eea  
	
	(ii) it is  a homogeneous function of $v^i$ of degree $n$,  
	\bea
	Q^{(n)}(z, {\mathfrak c} \,v)={\mathfrak c}^n\,Q^{(n)}(z,v)~, \qquad {\mathfrak c} \in \mathbb{C}\setminus  \{0\}~;
	\label{weight}
	\eea	
	
	(iii) it possesses the  superconformal transformation law
	\begin{align}
		&\qquad \qquad\quad\d_\cK Q^{(n)} 
		= \Big( \x^A \de_A + \L^{ij} J_{ij}+\S\mathbb{D} \Big) Q^{(n)} ~,  
		\\ 
		J_{ij}  Q^{(n)}&= -\Big(v_{(i}v_{j)}{\pa}^{(-2)} 
		-\frac{n}{(v,u)}v_{(i}u_{j)}\Big) Q^{(n)}~, \quad \pa^{(-2)} := \frac{1}{(v,u)} u^{i} \frac{\pa}{\pa v^{i}}~.
	\end{align}
\end{subequations}
By construction, $Q^{(n)}$ is independent of the isotwistor $u^i$, 
\bea
\pa^{(2)}   Q^{(n)}  =0~,\qquad
\pa^{(2)} := (v,u) v^{i} \frac{\pa}{\pa u^{i}} ~.
\eea
The variation $\d_\cK Q^{(n)} $ is characterised by 
the same property, $\pa^{(2)}\d_\cK Q^{(n)} =0$,
due to the homogeneity condition (\ref{weight}). 

Consistency of the analyticity conditions \eqref{ana} with the superconformal algebra \eqref{N=2ConfAlg} leads to non-trivial implications. Specifically, its dilatation weight is constrained by
\begin{align}
	\mathbb{D} Q^{(n)} = n Q^{(n)} ~.
\end{align}
To prove this, it is useful to make use of the identities
\begin{subequations}
	\label{SIdentities}
	\begin{align}
		\big \{ S_{\a}^{(1)} , \nabla_\b^{(1)} \big \} &= 4 \ve_{\a \b} \partial^{(2)} ~, \quad \big \{S_\a^{(1)}, \nabla_\b^{(-1)} \big \} = - \ve_{\a \b} \Big(2 \mathbb{D} - \mathbb{Y} + 2 \partial^{(0)} \Big) + 4 M_{\a \b}~, \\
		\big \{ S_\a^{(-1)} , \nabla_\b^{(1)} \big \} &= \ve_{\a \b} \Big( 2 \mathbb{D} - \mathbb{Y} - 2 \partial^{(0)} \Big) - 4 M_{\a \b}~, \quad \big \{ S_\a^{(-1)} , \nabla_\b^{(-1)} \big \} = - 4 \ve_{\a \b} \partial^{(-2)}~, \\
		\big \{ \bar{S}_{\ad}^{(1)} , \bar{\nabla}_\bd^{(1)} \big \} &= 4 \ve_{\ad \bd} \partial^{(2)} ~, \quad \big \{\bar{S}_\ad^{(1)}, \bar{\nabla}_\bd^{(-1)} \big \} = - \ve_{\ad \bd} \Big(2 \mathbb{D} + \mathbb{Y} + 2 \partial^{(0)} \Big) + 4 \bar{M}_{\ad \bd}~, \\
		\big \{ \bar{S}_\ad^{(-1)} , \bar{\nabla}_\bd^{(1)} \big \} &= \ve_{\ad \bd} \Big( 2 \mathbb{D} + \mathbb{Y} - 2 \partial^{(0)} \Big) - 4 \bar{M}_{\ad \bd}~, \quad \big \{ \bar{S}_\ad^{(-1)} , \bar{\nabla}_\bd^{(-1)} \big \} = - 4 \ve_{\ad \bd} \partial^{(-2)}~,
	\end{align}
\end{subequations}
which are valid when acting on any superfield $U(z,v,u)$ (with suppressed Lorentz indices and weight).
Here we have made the definitions
\begin{subequations}
	\begin{align}
		S^{(1)}_\a &= v_i S_\a^i~, \quad S^{(-1)}_\a = \frac{1}{(v,u)} u_i S_\a^i ~, \quad
		\bar{S}^{(1)}_\ad = v_i \bar{S}_\ad^i~, \quad \bar{S}^{(-1)}_\ad = \frac{1}{(v,u)} u_i \bar{S}_\ad^i ~, \\
		& \qquad \qquad\quad\nabla^{(-1)}_\a = \frac{1}{(v,u)} u_i \nabla_\a^i ~, \qquad \bar{\nabla}^{(-1)}_\ad = \frac{1}{(v,u)} u_i \bar{\nabla}_\ad^i ~.
	\end{align}
\end{subequations}

There exists 
a real structure on the space of projective multiplets \cite{Rosly,GIKOS,LR1}.
Given a  weight-$n$ projective multiplet $ Q^{(n)} (v^{i})$, 
its {\it smile conjugate} $ \breve{Q}^{(n)} (v^{i})$ is defined via 
\bea
Q^{(n)}(v^{i}) \longrightarrow  {\bar Q}^{(n)} ({\bar v}_i) 
\longrightarrow  {\bar Q}^{(n)} \big({\bar v}_i \to -v_i  \big) =:\breve{Q}^{(n)}(v^{i})~,
\label{smile-iso}
\eea
with ${\bar Q}^{(n)} ({\bar v}_i)  :=\overline{ Q^{(n)}(v^{i} )}$
the complex conjugate of  $ Q^{(n)} (v^{i})$, and ${\bar v}_i$ the complex conjugate of 
$v^{i}$. It may be shown that $ \breve{Q}^{(n)} (v)$ is a weight-$n$ projective multiplet.
In particular, unlike the complex conjugate of $Q^{(n)}(v) $, the superfield  $ \breve{Q}^{(n)} (v)$
obeys the analyticity constraints \eqref{ana}.
One can also check that 
\bea
\breve{ \breve{Q}}^{(n)}(v) =(-1)^n {Q}^{(n)}(v)~.
\label{smile-iso2}
\eea
Therefore, for even $n$, one can define real projective multiplets, 
which are constrained by $\breve{Q}^{(2n)} = {Q}^{(2n)}$.
Note that geometrically, the smile-conjugation is complex conjugation composed
with the antipodal map on the projective space ${\mathbb C}P^1$.

The isotwistor variables $v^i$ are homogeneous coordinates for ${\mathbb C}P^1$. It is often useful to deal with an inhomogeneous complex coordinate that can be introduced on an open domain of  ${\mathbb C}P^1$ obtained by removing a single point.  
We identify the north chart of ${\mathbb C}P^1$ with the open subset consisting of those points for which 
the first component  of $v^i = (v^{\1}, v^{\2})$ is non-zero,  $v^{\1} \neq 0$.
The north chart of ${\mathbb C}P^1$ may be parametrised by the complex coordinate 
$\z= v^{\2}/v^{\1} \in \mathbb C$. The only point of ${\mathbb C}P^1 $ outside the north 
chart is characterised by $v_\infty^i = (0, v^{\2})$ and describes an infinitely separated point.
Given a weight-$n$ projective multiplet $ Q^{(n)}(v^i)$, we can associate with it 
a rescaled superfield
\bea
 Q^{[n]}(\z) \propto Q^{(n)}(v^i)~, \qquad 
\frac{\pa}{\pa \bar \z} Q^{[n]} =0 ~.
\eea
The explicit form of $ Q^{[n]}(\z) $ depends on the multiplet under consideration. 
For example, it is given by the relations \eqref {arctic1} and \eqref{antarctic1} for the arctic 
${\U}^{(1)}(v)$ and antarctic  $\breve{\U}^{(1)}(v)$ weight-1 multiplets, respectively. 

The south chart of ${\mathbb C}P^1$ is defined to consist of those points for which 
the second component  of $v^i = (v^{\1}, v^{\2})$ is non-zero,  $v^{\2} \neq 0$.
It is naturally parametrised by $1/\z$. Since the projective action principle
\eqref{InvarAc} involves only a contour integral in ${\mathbb C}P^1$, it suffices to work in the north chart, for a point outside of the integration contour  $\g$ may be identified with the north pole. 

\subsection{Superconformal isotwistor multiplets}

There is a simple construction to generate covariant projective multiplets which makes use of so-called isotwistor superfields \cite{KLRT-M1, Kuzenko:2011xg}. 
By definition, a {\it superconformal isotwistor multiplet} of weight $n$,  $U^{(n)} (z,v)$, is a primary tensor superfield (with suppressed Lorentz indices)
which is holomorphic with respect to 
the isospinor variables $v^i $ on an open domain of 
${\mathbb C}^2 \setminus  \{0\}$ and has the following properties:

\begin{subequations}
	(i) it is a homogeneous function of $v^i$ of degree $n$,
	\bea
	&&U^{(n)}(z,{\mathfrak c} \,v)\,=\,{\mathfrak c}^n\,U^{(n)}(z,v)~, \qquad {\mathfrak c}\in 
	{\mathbb C} \setminus  \{0\}
	~;
	\eea
	
	(ii) it is  characterised by the gauge transformation law
	\bea
	\d_\cK U^{(n)} 
	&=& \Big( \x^A \de_A
	+  \hf \L^{ab} M_{ab}
	+\L^{ij} J_{ij} 
	+\S\mathbb{D}
	\Big) 
	U^{(n)} ~,  ~~~
	\non \\
	J_{ij}  U^{(n)}&=& -\Big(v_{(i}v_{j)}{\pa}^{(-2)} 
	-\frac{n}{(v,u)}v_{(i}u_{j)}\Big) U^{(n)}
	~.
	\label{iso2}
	\eea 
\end{subequations}
Now, given a Lorentz-scalar isotwistor superfield $U^{(n-4)}$ of dimension $n-2$ 
\bea
\mathbb{D}  U^{(n-4)}=
(n-2)  U^{(n-4)}~.
\label{iso3}
\eea
The following descendant 
\bea
Q^{(n)}=\nabla^{(4)}U^{(n-4)}
\eea
is a covariant projective multiplet.  Here we have introduced the operators
\bea 
\nabla^{(4)} = \frac{1}{16} \nabla^{(2)} \bar \nabla^{(2)}~, \qquad
\nabla^{(2)} = v_i v_j \nabla^{ij} ~, \quad \bar \nabla^{(2)} = v_i v_j \bar \nabla^{ij} ~.
\label{aprojector}
\eea
Isotwistor superfields will play an important role in what follows.

\subsection{Superconformal action} 

We recall that the projective action principle in conformal superspace is formulated in terms of 
a Lagrangian $\cL^{(2)}$ which is a real superconformal weight-2 projective multiplet. 
The locally superconformal action functional  is given by
\bea
S&=&
\frac{1}{2\pi} \oint_\g (v, \rd v)
\int \rd^{4|8}z\, E\, \frac{ X}{ \nabla^{(4)} X }
\cL^{(2)}~, \qquad  (v, \rd v) := v^i \rd v_i ~,
\label{InvarAc}
\eea
where $\g$ denotes a closed integration contour, 
$\rd^{4|8}z=\rd^4x\,\rd^4\q \rd^4 \bar \q$ is the full superspace integration measure and
$E^{-1} = {\rm Ber}(E_A{}^M)$. Finally,  the fourth-order operator $ \nabla^{(4)} $ is defined in \eqref{aprojector},
and $X (v)$ is a superconformal weight-$0$ isotwistor multiplet of dimension $+2$
The action functional may be shown to be independent of $X$.\footnote{Actually, $X$ in \eqref{InvarAc} may be replaced with a superconformal weight-$n$ isotwistor multiplet
$X^{(n)}$ of dimension $n+2$, see  \cite{KRTM} for more details.}
The action can be reduced to components to result with 
\cite{KTM08, Butter:2014gha}
\bea
S =  \frac{1}{2\p} \oint_{\g}  (v, \rd v)
\int {\rm d}^4x \, e \, \left\{  \nabla^{(-4)} \cL^{(2)}\Big| +\cdots \right\} ~, 
\label{PAP}
\eea
where 
$\nabla^{(-4)}$ is the following differential operator
\bea
\nabla^{(-4)} := \frac{1}{16} (\nabla^{(-1) })^2  (\bar \nabla^{(-1)})^2~, 
\eea
and we have made use of the notational shorthand $\F | \equiv \F |_{\q_i = \bar {\q}^i = 0}$. In the component action \eqref{PAP}, the ellipsis denotes those terms which contain factors of $\nabla^{(-1)}_\a $ and $\bar \nabla^{(-1)}_{\ad} $
of third and lower orders multiplied by functions of the supergravity fields. 
As usual, $e  = \det (e_m{}^a)$,  where $e^a = \rd x^m e_m{}^a (x) $ is the spacetime vierbein. 
By construction, the action is independent of $u^i$.

\subsection{Superconformal polar hypermultiplet}

We recall that an off-shell polar hypermultiplet is described in terms of an arctic weight-1 multiplet 
$\U^{(1)}(v)$ and its smile-conjugate antarctic multiplet $\breve{\U}^{(1)} (v)$.
By definition, the off-shell arctic weight-$1$ multiplet, $\U^{(1)}(v)$, is a superconformal projective multiplet
which is holomorphic in the north chart of ${\mathbb C}P^1$
\bea
\U^{(1)} ( v) &=&  v^{\1}\, \U ( \z) ~, \qquad 
\U ( \z) = \sum_{k=0}^{\infty} \U_k  \z^k 
~. 
\label{arctic1}
\eea
The smile-conjugate of $\U^{(1)}$ is an antarctic multiplet, $\breve{\U}^{(1)}(v)$, which is
holomorphic in the south chart of ${\mathbb C}P^1$
\bea
\breve{\U}^{(1)} (v) &=& v^{\2}  \, \breve{\U}(\z) =
v^{\1} \,\z \, \breve{\U} (\z) ~, \quad
\breve{\U}( \z) = \sum_{k=0}^{\infty}  {\bar \U}_k \,
\frac{(-1)^k}{\z^k}~.
\label{antarctic1}
\eea

The action for a free superconformal hypermultiplet is then given by
\bea
\label{HM}
S_\text{HM}=  \frac{\ri}{2\p} \oint_{\g}  (v, \rd v)
\int \rd^{4|8}z\, E\, \frac{ X}{ \nabla^{(4)} X } \, \breve{\U}^{(1)} \U^{(1)} 
~.
\eea
Varying 
this 
action 
yields the equation of motion
\begin{align}
	\label{HMEoM}
	(\partial^{(-2)})^2\U^{(1)} ( v) = 0 ~,
\end{align}
and hence the on-shell hypermultiplet is given by
\begin{align}
	\U^{(1)} ( v) = v_i \U^i~,
\end{align}
where $\U^i$ is a primary isospinor. Keeping in mind the analyticity conditions \eqref{ana}, we find that it is subject to the constraints
\bea
\label{FSHM}
\nabla_{\a}^{(i} \U^{j)} = 0~, \qquad \bar \nabla_{\ad}^{(i} \U^{j)} = 0
\quad \implies \quad \Big( \nabla^a \nabla_a +\frac 18 B\Big)\U^i = 0~,
\eea
which define the on-shell Fayet-Sohnius hypermultiplet \cite{Fayet,Sohnius78} coupled to the Weyl multiplet for conformal supergravity.

\section{Isotwistor conformal supercurrents} \label{IsotwistorCSCs}

Let us start by recalling 
the conformal supercurrent multiplets introduced in \cite{KR21}. 
Given positive integers $m$ and $n$, a conformal supercurrent multiplet $J^{\a(m) \ad(n)}$
is a primary tensor superfield obeying the constraints
\begin{subequations}
\label{SuperC1}
\bea
\nabla_\b^i J^{\b \a(m-1) \ad(n)} &=& 0 \quad \Longrightarrow \quad \nabla^{ij} J^{\a(m) \ad(n)} = 0 ~, \\
\bar{\nabla}_\bd^i J^{\a(m) \bd \ad(n-1)} &=& 0 \quad \Longrightarrow \quad \bar{\nabla}^{ij} J^{\a(m) \ad(n)} = 0 ~.
\eea
\end{subequations}
These constraints uniquely fix the superconformal properties of $J^{\a(m) \ad(n)}$
\bea
\mathbb{D} J^{\a(m) \ad(n)} = \hf (m+n+4) J^{\a(m) \ad(n)} ~, \quad \mathbb{Y} J^{\a(m) \ad(n)} = -(m-n) J^{\a(m) \ad(n)} ~.
\eea
For $m=n=s$, $J^{\a(s) \ad(s)}$ is invariant under $\sU(1)_R$ transformations and thus 
$J^{\a(s) \ad(s)}$ is restricted 
 to be real. This special case was first described in Minkowski superspace in \cite{HST} without discussing the superconformal properties of $J^{\a(s) \ad(s)}$.
 
In the $n = 0$ case, the constraints \eqref{SuperC1} are replaced with
\begin{subequations}
\label{SuperC2}
\bea
\nabla_\b^i J^{\b \a(m-1)} &=& 0 \quad \Longrightarrow \quad \nabla^{ij} J^{\a(m) } = 0 ~, \\
\bar{\nabla}^{ij} J^{\a(m)} &=& 0 ~.
\eea
\end{subequations}
Consistency of \eqref{SuperC2} with the superconformal algebra implies:
\bea
\mathbb{D} J^{\a(m)} = \hf (m + 4) J^{\a(m)} ~, \quad \mathbb{Y} J^{\a(m)} = -m J^{\a(m)} ~.
\eea
Finally, if $m=n=0$, the supercurrent $J = \bar J$ obeys the constraints 
\bea\label{SuperC3}
\nabla^{ij} J = 0 ~, \quad \bar{\nabla}^{ij} J = 0 ~,
\eea
which  imply $\mathbb{D} J = 2 J $.
Constraints \eqref{SuperC3} define
the usual conformal supercurrent
\cite{HST,Sohnius,KT}.

For the on-shell hypermultiplet, which satisfies the equations of motion \eqref{FSHM}, the conserved conformal supercurrents $J^{\a(s) \ad(s)} $
were constructed in our previous work \cite{KR21}.
In terms of  $\U^i$ and its conjugate $\bar{\U}_i$, they are given 
by the following primary descendants:
\bea
\label{HMSC}
J^{\a(s) \ad(s)} &=& - \frac{\ri^{s}}{2} \sum^s_{k=0} (-1)^k {s \choose k}^2 (\nabla^{\aa})^k \U^i (\nabla^{\aa})^{s-k}\bar{\U}_i   \non \\
&& + \frac{\ri^{s+1}} {16} \sum_{k=0}^{s-1} (-1)^k {s \choose k} {s \choose k+1} \bigg\{ (\nabla^{\aa})^k \nabla^{\a i} \U_i (\nabla^{\aa})^{s-k-1}  \bar \nabla^{\ad j} \bar \U_j \non \\
&& \qquad \qquad \qquad \qquad \qquad \qquad - (\nabla^{\aa})^k \bar \nabla^{\ad i} \U_i (\nabla^{\aa})^{s-k-1} \nabla^{\a j}
\bar \U_j
\bigg\} ~,
\eea 
It should be noted that the $s=0$ case has been studied earlier in \cite{KT}.

Now, we present 
 off-shell extensions, $\cJ^{\a(s)\ad(s)}$,  
of the conserved supercurrents \eqref{HMSC}, which are constructed in terms of the off-shell polar hypermultiplet. 
We postulate that: (i)  $\cJ^{\a(s)\ad(s)}$ is a real primary isotwistor superfield of dimension $s+2$
\begin{align}
	\partial^{(2)} \cJ_{\a(s) \ad(s)}&= 0~, \qquad ~~ \breve{\cJ}_{\a(s) \ad(s)} =\cJ_{\a(s) \ad(s)}~, \non \\
	K^B \cJ_{\a(s) \ad(s)}&= 0 ~, \qquad \mathbb{D} \cJ_{\a(s) \ad(s)} = (s+2)\cJ_{\a(s) \ad(s)}~;
\end{align}
and (ii) $\cJ^{\a(s)\ad(s)}$ reduces to \eqref{HMSC} on-shell,
\begin{align}
	\cJ^{\a(s) \ad(s)} \xrightarrow{\U^{(1)} ( v)\,=\, v_i \U^i}
	J^{\a(s) \ad(s)}~.
\end{align}
It turns out that these conditions have the unique solution
\begin{align}
	\label{IsoSC}
	\cJ^{\a(s) \ad(s)} &= \frac{\ri^s}{2} \sum_{k=0}^s (-1)^k \binom{s}{k}^2 \bigg \{ \frac{1}{k+1} (\nabla^{\aa})^k \partial^{(-2)} \breve{\U}^{(1)} (\nabla^\aa)^{s-k} \U^{(1)} \non \\
	& \qquad\qquad - \frac{1}{s-k+1}(\nabla^{\aa})^k \breve{\U}^{(1)} (\nabla^\aa)^{s-k} \partial^{(-2)} \U^{(1)} \non \\
	& \qquad\qquad - \frac{\ri}{2} \frac{(s-k)^2}{(k+1)(k+2)} (\nabla^\aa)^k \nabla^{\a(-1)} \nabla^{\ad(-1)} \breve{\U}^{(1)} (\nabla^\aa)^{s-k-1} \U^{(1)} \non \\
	& \qquad\qquad - \frac{\ri}{2} \frac{s-k}{s-k+1} (\nabla^\aa)^k \breve{\U}^{(1)} (\nabla^\aa)^{s-k-1} \nabla^{\a(-1)} \bar{\nabla}^{\ad(-1)} \U^{(1)} \non \\
	& \qquad\qquad + \frac{\ri}{2} \frac{s-k}{k+1} (\nabla^{\aa})^k \nabla^{\a(-1)} \breve{\U}^{(1)} (\nabla^\aa)^{s-k-1} \bar{\nabla}^{\ad (-1)}\U^{(1)}  \non \\
	& \qquad\qquad - \frac{\ri}{2} \frac{s-k}{k+1} (\nabla^{\aa})^k \bar{\nabla}^{\ad (-1)} \breve{\U}^{(1)} (\nabla^\aa)^{s-k-1} \nabla^{\a(-1)} \U^{(1)} \bigg \}~.
\end{align}
In what follows we will refer to $\cJ^{\a(s)\ad(s)}$ as {\it isotwistor conformal supercurrents}. Below, we will show how the same supercurrents arise from the Noether procedure.

\section{Hypermultiplet coupled to SCHS multiplets}
\label{Section3}

In this section we employ a superfield Noether procedure, see e.g. \cite{KPR} for a general discussion, to engineer manifestly superconformal interactions between a polar hypermultiplet and an infinite tower of superconformal (higher-spin) gauge multiplets.

\subsection{Local transformations of polar hypermultiplet}
\label{Section3.1}

To perform the Noether procedure, it is first necessary to construct possible local transformation rules for $\U^{(1)}$ which are consistent with its kinematic properties; they should preserve the space of arctic weight-1 multiplets. The appropriate transformations were sketched in \cite{KR23} and are of the form\footnote{Actually, the operators $\cU_{[0]}$ and $\cU_{[1]}$ were computed in \cite{KR23}.}
\begin{align}
	\label{HMGT}
	\d_{\O} \U^{(1)} = - \cU \U^{(1)} = -\sum_{s=0}^{\infty} \cU_{[s]} \U^{(1)}~, \quad
	\cU_{[s]} \propto \nabla^{(4)} \Big \{ \O^{\a(s) \ad(s) (-2)} (\nabla_{\aa})^{s} \partial^{(-2)} + \dots \Big \}~,
\end{align}
where $\O^{\a(s) \ad(s)(-2) }$ is a primary isotwistor superfield of dimension $-(s+2)$
\begin{align}
	K^B \O^{\a(s) \ad(s)(-2) } = 0~, \qquad \mathbb{D} \O^{\a(s) \ad(s)(-2) } = -(s+2) \O^{\a(s) \ad(s)(-2) } ~,
\end{align}
and the ellipses denote additional terms necessary to preserve the kinematics of $\U^{(1)}$. Specifically, one must impose the conditions
\begin{subequations}
\begin{align}
	\mathbb{D} \cU_{[s]} \U^{(1)} &= \partial^{(0)} \cU_{[s]} \U^{(1)} = \cU_{[s]} \U^{(1)} ~, \qquad \partial^{(2)} \cU_{[s]} \U^{(1)} = 0 ~, \qquad K^A \cU_{[s]} \U^{(1)} = 0~, \\
	\nabla_\a^{(1)} &\cU_{[s]} \U^{(1)} = 0 ~, \qquad \bar{\nabla}_\ad^{(1)} \cU_{[s]} \U^{(1)} = 0~, \qquad \cU_{[s]} \U^{(1)}(v) = v^{\underline 1}\sum_{k=0}^{\infty} (\cU_{[s]} \U)_k \z^k~.
\end{align}
\end{subequations}
Direct calculations lead to the unique solution (modulo overall normalisation)
\begin{align}
	\label{3.4}
	\cU_{[s]} &= - \frac{\ri^{s}}{2(s+1)} \nabla^{(4)} \bigg \{ \sum_{k=0}^s 
	\binom{s}{k} \bigg \{ \binom{2s-k+2}{s-k+1} (\nabla_\aa)^k \O^{\a(s) \ad(s) (-2)} (\nabla_\aa)^{s-k} \partial^{(-2)} \non \\
	& \qquad + \binom{2s-k+1}{s-k} (\nabla_\aa)^k \partial^{(-2)} \O^{\a(s) \ad(s) (-2)} (\nabla_\aa)^{s-k} \non \\
	& \qquad + \frac{\ri}{2} (s-k) \binom{2s-k+2}{s-k+1} (\nabla_\aa)^k \O^{\a(s) \ad(s) (-2)} (\nabla_\aa)^{s-k-1} \nabla_\a^{(-1)} \bar{\nabla}_\ad^{(-1)} \non \\
	& \qquad + \frac{\ri}{2} (s-k) \binom{2s-k}{s-k-1} \nabla_\a^{(-1)} \bar{\nabla}_\ad^{(-1)} (\nabla_\aa)^k \O^{\a(s) \ad(s) (-2)} (\nabla_\aa)^{s-k-1} \non \\
	& \qquad + \frac{\ri}{2} (s-k) \binom{2s-k+1}{s-k} \nabla_\a^{(-1)} (\nabla_\aa)^k \O^{\a(s) \ad(s) (-2)} (\nabla_\aa)^{s-k-1} \bar{\nabla}_\ad^{(-1)} \non \\
	& \qquad - \frac{\ri}{2} (s-k) \binom{2s-k+1}{s-k} \bar{\nabla}_\ad^{(-1)} (\nabla_\aa)^k \O^{\a(s) \ad(s) (-2)} (\nabla_\aa)^{s-k-1} \nabla_\a^{(-1)} \bigg\} \bigg\}~,	
\end{align}
where the gauge parameter $\O^{\a(s) \ad(s) (-2)}$ is holomorphic in the north chart of $\mathbb{C}P^1$ in the sense that it may be expressed as a Taylor series in $\z$.

It should be noted that \eqref{HMGT} is not the most general transformation preserving the off-shell properties of $\U^{(1)}$. In particular, one could consider 
\begin{align}
	\d_{\O} \U^{(1)} = \ri \O \U^{(1)}~, \qquad \nabla^{(1)}_\a \O = 0 ~, \quad \bar{\nabla}^{(1)} _\ad\O =0~, \quad \partial^{(2)} \O =0~,
\end{align}
where $\O$ is a weight-$0$ arctic multiplet. Gauging this symmetry yields the well-known interactions with an Abelian vector multiplet, see e.g. \cite{KRTM} for a review. Additionally, we have ignored transformations with contributions vanishing on-shell \eqref{HMEoM}. These turn out to be necessary for closure of the gauge algebra\footnote{See \cite{KPR} for a discussion of the analogous transformations in $\cN=0$ and $\cN=1$ cases.} and will be discussed in section \ref{Section4.1}.

\subsection{Cubic Noether couplings} 
\label{Section3.2}

Having constructed the appropriate transformation laws for a polar hypermultiplet above, it is now necessary to deform the free theory \eqref{HM} such that the resulting model remains superconformal and is invariant under \eqref{HMGT}. To this end, we propose the model
\begin{align}
	\label{DeformedAction}
	S_\text{Cubic} =  \frac{\ri}{2\p} \oint_{\g}  (v, \rd v)
	\int {\rm d}^{4|8}z \, E \, \frac{ X}{ \nabla^{(4)} X } \big( \breve{\U}^{(1)} (1 + \cH) \U^{(1)} \big) ~.
\end{align}
Here $\cH$ is a 
differential operator with the following properties:

(i) it is Hermitian in the sense that it coincides with its adjoint, $\cH = \cH^{\dagger}$, where we define the adjoint $\cV^{\dagger}$ of a differential operator $\cV$ on the space of polar hypermultiplets by
\begin{subequations} 
	\label{HDDMConditions}
	\begin{align}
		\frac{\ri}{2\p} \oint_{\g}  (v, \rd v)
		\int {\rm d}^{4|8}z \, E \, \frac{ X}{ \nabla^{(4)} X } \big( \widebreve{(\cV^{\dagger} {\U}^{(1)}}) \U^{(1)} \big) =
		\frac{\ri}{2\p} \oint_{\g}  (v, \rd v)
		\int {\rm d}^{4|8}z \, E \, \frac{ X}{ \nabla^{(4)} X } \big( \breve{\U}^{(1)} \cV \U^{(1)} \big) ~;
	\end{align}
	(ii) it preserves the superconformal and kinematic properties of polar multiplets of unit weight
		\begin{align}
			\mathbb{D} \cH \U^{(1)} &= \partial^{(0)} \cH \U^{(1)} = \cH \U^{(1)} ~, \qquad \partial^{(2)} \cH \U^{(1)} = 0 ~, \qquad K^A \cH \U^{(1)} = 0~, \\
			& \qquad \qquad \quad \nabla_\a^{(1)} \cH \U^{(1)} = 0 ~, \qquad \bar{\nabla}_\ad^{(1)} \cH \U^{(1)} = 0~;
		\end{align}
	(iii) it is defined modulo the gauge transformations
		\begin{align}
			\label{GT}
			\d_\O \cH = {\cU}^{\dagger} (1+\cH) + (1+\cH) \cU~.
		\end{align}
\end{subequations}

A calculation similar to the one undertaken in 
the previous 
section 
allows one to uniquely determine (modulo overall real coefficient) the form of $\cH$. Specifically,
\begin{subequations}
	\label{HDDM}
	\begin{align}
	\cH = \sum_{s=0}^\infty \cH_{[s]} ~, 
\end{align}
where the operator $\cH_{[s]}$ is given by
\begin{align}
	\cH_{[s]} &= \frac{\ri^{s+1}}{2(s+1)} \nabla^{(4)} \bigg \{ \sum_{k=0}^s 
	\binom{s}{k} \bigg \{ \binom{2s-k+2}{s-k+1} (\nabla_\aa)^k \cH^{\a(s) \ad(s) (-2)} (\nabla_\aa)^{s-k} \partial^{(-2)} \non \\
	& \qquad + \binom{2s-k+1}{s-k} (\nabla_\aa)^k \partial^{(-2)} \cH^{\a(s) \ad(s) (-2)} (\nabla_\aa)^{s-k} \non \\
	& \qquad + \frac{\ri}{2} (s-k) \binom{2s-k+2}{s-k+1} (\nabla_\aa)^k \cH^{\a(s) \ad(s) (-2)} (\nabla_\aa)^{s-k-1} \nabla_\a^{(-1)} \bar{\nabla}_\ad^{(-1)} \non \\
	& \qquad + \frac{\ri}{2} (s-k) \binom{2s-k}{s-k-1} \nabla_\a^{(-1)} \bar{\nabla}_\ad^{(-1)} (\nabla_\aa)^k \cH^{\a(s) \ad(s) (-2)} (\nabla_\aa)^{s-k-1} \non \\
	& \qquad + \frac{\ri}{2} (s-k) \binom{2s-k+1}{s-k} \nabla_\a^{(-1)} (\nabla_\aa)^k \cH^{\a(s) \ad(s) (-2)} (\nabla_\aa)^{s-k-1} \bar{\nabla}_\ad^{(-1)} \non \\
	& \qquad - \frac{\ri}{2} (s-k) \binom{2s-k+1}{s-k} \bar{\nabla}_\ad^{(-1)} (\nabla_\aa)^k \cH^{\a(s) \ad(s) (-2)} (\nabla_\aa)^{s-k-1} \nabla_\a^{(-1)} \bigg\} \bigg \}~.
\end{align}
\end{subequations}
Here $\cH^{\a(s) \ad(s) (-2)} = \breve{\cH}^{\a(s) \ad(s) (-2)}$ is a primary isotwistor superfield of dimension $-(s+2)$
\begin{align}
	K^{B} \cH^{\a(s) \ad(s) (-2)} = 0~, \qquad \mathbb{D} \cH^{\a(s) \ad(s) (-2)} = -(s+2) \cH^{\a(s) \ad(s) (-2)}~.
\end{align}

Remarkably, by making use of eq. \eqref{HDDM}, the deformed action \eqref{DeformedAction} may be equivalently expressed as
\begin{align}
	\label{NC}
	S_\text{Cubic} &= S_\text{HM} + \frac{1}{2\p} \sum_{s=0}^{\infty}  \oint_{\g}  (v, \rd v)
	\int {\rm d}^{4|8}z \, E \,  \cH_{\a(s) \ad(s)}^{(-2)} \cJ^{\a(s) \ad(s)}
	~, 
\end{align}
where $\cJ^{\a(s) \ad(s)}$ are exactly the isotwistor conformal supercurrents \eqref{IsoSC}. Now, putting the hypermultiplet on-shell, and making use of the $v$-independence of $J^{\a(s)\ad(s)}$, we can make the identification
\begin{align}
	\label{2.14}
	H_{\a(s) \ad(s)} = \frac{1}{2\pi}\oint_{\g}  (v, \rd v) \, \cH_{\a(s) \ad(s)}^{(-2)}~,
\end{align}
where $H_{\a(s) \ad(s)}$ are the known superconformal gauge multiplets \eqref{A.5}. Thus, we will refer to $\cH_{\a(s) \ad(s)}^{(-2)}$ as {\it isotwistor prepotentials}.

It remains to determine how the prepotentials $\cH_{\a(s) \ad(s)}^{(-2)}$ transform under the gauge transformations \eqref{GT}. A routine calculation leads to
\begin{align}
	\label{IsotwistorGaugeSymmetry}
	\d_{\O} \cH_{\a(s) \ad(s)}^{(-2)} = \ri \Big(\O_{\a(s) \ad(s)}^{(-2)} - \breve{\O}_{\a(s) \ad(s)}^{(-2)} \Big) + \cO(\cH)~.
\end{align}
We note that gauge-invariance (to leading order) of the Noether coupling in eq. \eqref{NC} under \eqref{IsotwistorGaugeSymmetry} immediately follows from the $v$-independence of $\cJ^{\a(s)\ad(s)}$ when the hypermultiplet is on-shell. Equivalently, we see from eq. \eqref{2.14} that, to leading order, $H_{\a(s)\ad(s)}$ is invariant under \eqref{IsotwistorGaugeSymmetry}. On the other hand, $H_{\a(s) \ad(s)}$ is known to possess the gauge freedom \eqref{SCHSgt}. In the following subsection, we will deduce its projective-superspace origin.

\subsection{Pre-gauge transformations}

The above discussion hints that the model \eqref{DeformedAction} possesses a second `hidden' gauge symmetry acting as the origin of \eqref{SCHSgt}. Inspired by the harmonic-superspace analysis of \cite{KT}, we propose the following pre-gauge transformations of $\cH_{\a(s) \ad(s)}^{(-2)}$
\begin{subequations}
	\label{PreGT}
	\bea
	\label{PreGT1}
	s > 0:&& \qquad ~~ \d_\l \cH_{\a(s) \ad(s)}^{(-2)} = \bar{\nabla}_{\ad}^{(1)} \l_{\a(s) \ad(s-1) }^{(-3)} + \text{s.c.} ~, \\
	\label{PreGT2}
	s = 0:&& \qquad \quad~~\,	\d_\l \cH^{(-2)} = (\bar{\nabla}^{(1)})^2 \l^{(-4)} + \text{s.c.} ~,
	\eea
\end{subequations}
where `$\text{s.c.}$' denotes smile conjugate, and the gauge parameters $\l$ are necessarily holomorphic in the intersection of the north and south charts of $\mathbb{C}P^1$. Remarkably, it may be shown that operator $\cH$, and hence the action \eqref{DeformedAction}, is inert under \eqref{PreGT}
\begin{align}
	\d_\l \cH = 0 \quad \implies \quad \d_\l S_\text{Cubic} = 0~.
\end{align} 

Now, keeping in mind the relationship between $\cH_{\a(s) \ad(s)}^{(-2)}$ and $H_{\a(s) \ad(s)}$ given in eq. \eqref{2.14}, we compute the parameters of \eqref{SCHSgt} in terms of those for \eqref{PreGT} 
\begin{subequations}
	\begin{align}
	s > 0:& \qquad \l_{\a(s) \ad(s-1) i} = \frac{1}{2\p} \oint_{\g}  (v, \rd v) \, v_i  \l_{\a(s) \ad(s-1) }^{(-3)} ~, \\
	s = 0:& \qquad \qquad \quad~\l_{ij} = \frac{1}{2\p} \oint_{\g}  (v, \rd v) \, v_i v_j  \l^{(-4)} ~.
	\end{align}
\end{subequations}
As a result, we have deduced the projective-superspace origin of the SCHS gauge transformations \eqref{SCHSgt}.

\subsection{Rigid symmetries of the free hypermultiplet action}

Having successfully engineered cubic couplings between an off-shell polar hypermultiplet and the isotwistor prepotentials $\cH_{\a(s) \ad(s)}^{(-2)}$ above, we now describe an interesting application of this construction. Specifically, we will directly read off an infinite class of rigid symmetries for the free theory \eqref{HM}. 

We begin by recalling that the $\O$-gauge transformations, defined by eq. \eqref{HMGT} and \eqref{IsotwistorGaugeSymmetry}, leave the cubic action \eqref{DeformedAction} invariant (to leading order in the prepotentials). Thus, switching off the prepotentials leads to
\begin{align}
	\d_\O S_\text{Cubic}\Big|_{\cH = 0} = \d_\O S_\text{HM} + \frac{1}{2\p} \sum_{s=0}^{\infty} \oint_{\g}  (v, \rd v)
	\int {\rm d}^{4|8}z \, E \,  \d_\O \cH_{\a(s) \ad(s)}^{(-2)} \cJ^{\a(s) \ad(s)}\Big|_{\cH = 0} = 0~.
\end{align}
Next, we note that the variation of the Noether coupling term above vanishes if $\O^{(-2)}_{\a(s)\ad(s)}$ is chosen to be smile-real\footnote{This implies that $\O^{(-2)}_{\a(s)\ad(s)}$ is holomorphic in the intersection of the north and south charts of $\mathbb{C}P^1$, which has highly non-trivial implications.}
\begin{align}
	\d_\O \cH_{\a(s) \ad(s)}^{(-2)} = \cO(\cH) \quad \Longleftrightarrow \quad \O^{(-2)}_{\a(s)\ad(s)} = \breve{\O}^{(-2)}_{\a(s)\ad(s)}~.
\end{align}
This implies the important result
\begin{align}
	\d_\O S_\text{HM} = 0 \quad \Longleftrightarrow \quad \d_\O \U^{(1)} = - \cU \U^{(1)}\Big|_{\O = \breve{\O}}~,
\end{align}
which defines an infinite class of rigid symmetries for the free theory \eqref{HM}.

As is well-known, the rigid symmetries of a given action also preserve its equations of motion. Higher symmetries for an on-shell hypermultiplet were computed in \cite{KR23}, and we now compare them with the rigid symmetries obtained above. We recall that the symmetries of \cite{KR23} take the form
\begin{align}
	\label{3.19}
	\d_\xi \U^{(1)} = - \sum_{s=1}^{\infty} \mathfrak{D}_{[s]} \U^{(1)}~, \qquad \mathfrak{D}_{[s]} \propto \xi^{\a(s) \ad(s)} (\nabla_{\aa})^s + \dots ~,
\end{align}
where the ellipses denote additional terms necessary to preserve both the superconformal properties and on-shell constraints of $\U^{(1)}$ while $\xi^{\a(s) \ad(s)} = \xi^{\a(s) \ad(s)}(z)$ is a conformal Killing tensor\footnote{The conformal Killing tensor superfields of Minkowski superspace were introduced in \cite{HL1}.}
\begin{align}
	\label{CKT}
	\nabla_\a^i \xi_{\a(s) \ad(s)} = 0 ~, \qquad \bar{\nabla}_{\ad}^i \xi_{\a(s) \ad(s)} = 0~.
\end{align}
On the other hand, one may rewrite the rigid symmetries of $\U^{(1)}$ in the form
\begin{align}
	\label{3.21}
	\d_\L \U^{(1)} = - \sum_{s=0}^{\infty} \cU_{[s]} \U^{(1)}~, \qquad \cU_{[s]} \propto \L^{\a(s+1) \ad(s+1)} (\nabla_{\aa})^{s+1} + \dots ~,
\end{align}
where we have introduced the isotwistor superfield $\L_{\a(s+1) \ad(s+1)}$
\begin{align}
	\L_{\a(s+1) \ad(s+1)} = \ri \nabla_\a^{(1)} \bar{\nabla}_\ad^{(1)} \O_{\a(s) \ad(s)} = \breve{\L}_{\a(s+1) \ad(s+1)}~, 
\end{align}
which is subject to the analyticity-like constraints
\begin{align}
	\label{3.23}
	\nabla_\a^{(1)} \L_{\a(s+1) \ad(s+1)} = 0 ~, \qquad \bar{\nabla}_\ad^{(1)} \L_{\a(s+1) \ad(s+1)} = 0~.
\end{align}
However, since the rigid symmetries \eqref{3.21} take the same functional form as the higher symmetries \eqref{3.19}, the two families are necessarily equivalent. In particular, $\L_{\a(s+1) \ad(s+1)}$ must be $v$-independent, and thus a conformal Killing tensor
\begin{align}
	\partial^{(-2)} \L_{\a(s+1) \ad(s+1)} = 0 \quad \implies \quad \nabla_\a^{i} \L_{\a(s+1) \ad(s+1)} = 0 ~, \qquad \bar{\nabla}_\ad^{i} \L_{\a(s+1) \ad(s+1)} = 0~.
\end{align}

\section{Discussion}
\label{Section4}

In closing, we emphasise that the superconformal higher-spin multiplets $H_{\a(s)\ad(s)}$ introduced\footnote{See also \cite{HST} where these prepotentials were mentioned without describing their superconformal properties.}
 in \cite{KR21}  have now been understood in terms of their unconstrained isotwistor prepotentials $\cH_{\a(s) \ad(s)}^{(-2)}$. While it was shown in \cite{KR21} that the former naturally appear in the linearised higher-spin super-Weyl tensors \eqref{HSSW}, and thus in the gauge-invariant action \eqref{SCHSaction}, the latter have proven to be necessary for engineering cubic couplings to an off-shell hypermultiplet. Hence, their features are complementary, and both constructions play a vital role. 

The remainder of this section is devoted to exploring extensions of the analysis undertaken above. In particular, we lay the grounds for future work and provide important connections to related results in the literature.

\subsection{Consistency to all orders}
\label{Section4.1}

As mentioned in section \ref{Section3.1}, the transformations \eqref{HMGT} are not the most general ones preserving the off-shell properties of $\U^{(1)}$ as we have ignored transformations containing contributions which vanish on-shell.
These arise naturally when studying the algebra of operators $\cU_{[s]}$ as their commutators involve terms proportional to $(\partial^{(-2)})^2$, which defines the equation of motion for $\U^{(1)}$, see eq. \eqref{HMEoM}. Thus, as argued in \cite{KPR} for the $\cN=0$ and $\cN=1$ supersymmetric cases, one must enlarge the algebra of gauge transformations by such structures to ensure closure (modulo trivial symmetries).

Keeping in mind the approach advocated in \cite{KPR}, 
this may be achieved by introducing the operator $\mathfrak{O}$, defined by
\begin{align}
	\mathfrak{O} \U^{(1)} := \nabla^{(4)} \bigg \{ \Xi (\partial^{(-2)})^2 \U^{(1)} \bigg \} ~,
\end{align}
which evidently annihilates $\U^{(1)}$ on-shell. To maintain the kinematics of $\U^{(1)}$, the compensating multiplet $\Xi(v)$ must be a primary isotwistor superfield of dimension $-2$,
\begin{align}
	K^B \Xi = 0 ~, \qquad \mathbb{D} \Xi = -2 \Xi~,
\end{align}
and holomorphic in the north chart of $\mathbb{C}P^1$. The local transformations for the polar hypermultiplet obtained in section \ref{Section3.1} can then be extended to
\begin{align}
	\label{closed}
	\d \U^{(1)} = - \bm{\cU} \U^{(1)} = - \sum_{s=0}^\infty \sum_{l=0}^{\lfloor s/2 \rfloor} \cU_{[s-2l]}^{[s,l]} \mathfrak{O}^l \U^{(1)}~,
\end{align}
where the operator $\cU^{[s,l]}_{[s-2l]}$ coincides with $\cU_{[s-2l]}$, see eq. \eqref{3.4}, with the sole exception that they are expressed in terms of different gauge parameters. We emphasise that the algebra of matter transformations \eqref{closed} closes.

In conjunction with this, it is necessary to extend the analysis undertaken in section \ref{Section3.2} and gauge these symmetries. Specifically, in accordance with \cite{KPR}, we associate with each transformation an `auxiliary' gauge prepotential\footnote{Such prepotentials were called `auxiliary' in \cite{KPR} as they describe purely gauge degrees of freedom; their gauge transformations should allow us to enforce the gauge $\cH_\text{aux} = 0$ at the expense of complicating the gauge transformations of the `physical' prepotentials $\cH^{(-2)}_{\a(s) \ad(s)}$ and matter multiplet.} $\cH_\text{aux}$ (with suppressed indices and weight) appearing in the primary operator $\bm{\cH}$, which extends $\cH$, see eq. \eqref{HDDM}, and is characterised by the gauge transformation law
\begin{align}
	\label{fullGT}
	\d \bm{\cH} = {\bm \cU}^{\dagger} (1+{\bm \cH}) + (1+{\bm \cH}) {\bm \cU}~.
\end{align}
The complete superconformal action for a polar hypermultiplet interacting with an infinite tower of SCHS multiplets is then given by
\begin{align}
	\label{4.1}
	\cS[\U,\breve{\U},\cH] =  \frac{\ri}{2\p} \oint_{\g}  (v, \rd v)
	\int {\rm d}^{4|8}z \, E \, \frac{ X}{ \nabla^{(4)} X } \big( \breve{\U}^{(1)} \re^{\bm \cH} \U^{(1)} \big) ~.
\end{align}
It is manifestly invariant under the finite gauge transformations
\begin{align}
	\U^{(1)} \longrightarrow \re^{-\bm{\cU}} \U^{(1)} ~, \qquad \re^{\bm \cH} \longrightarrow \re^{{\bm \cU}^\dagger} \re^{\bm \cH} \re^{\bm \cU}~,
\end{align}
which reduce to eq. \eqref{closed} and \eqref{fullGT} upon linearisation. We hope to fill in the details and complete this story in the future.

The above discussion has sketched the $\cN=2$ extension of the formalism developed in \cite{KPR} in the bosonic ($\cN=0$) and $\cN=1$ supersymmetric cases. The powerful feature of the approach of \cite{KPR} is that (super)conformal symmetry is manifestly realised. Its technical drawback is that one has to take care of  the inclusion of auxiliary gauge potentials required to close the gauge algebra. In the non-supersymmetric case, an alternative approach to the construction of an interacting conformal higher-spin theory has been developed \cite{BJM}.  
Specifically, these authors considered the following local transformation law of a complex scalar field $\varphi$
\begin{align}
	\d_\z \varphi = - \sum_{s=0}^\infty \cU_{[s]} \varphi ~, \qquad \cU_{[s]} = \z^{a(s)} (\partial_a)^s~,
	\label{4.7TL}
\end{align}
where the parameter $\z^{a(s)}$ is totally symmetric and traceful. The specific feature of this approach is that  %
the gauge algebra is manifestly closed. However, since the gauge parameters are traceful, the corresponding gauge fields inherit this property,\footnote{Their traceful components play the same role as the `auxiliary' gauge fields of \cite{KPR}.} which means neither can be primary. This 
means that 
conformal invariance of the theory is not manifest, see \cite{KPR} for more details.\footnote{It is worth pointing out that one may have conformal invariance without the fields being primary. This is the case e.g. in Metsaev's ordinary derivative formulation of the free bosonic CHS actions \cite{Metsaev2, Metsaev}.} 

It is not difficult to derive an $\cN=1$ generalisation of \eqref{4.7TL}.
Specifically, the appropriate local transformation law of a chiral scalar $\F$
is given by:
\begin{subequations}
	\bea
	&& \qquad \d_\z \F = - \sum_{s=0}^{\infty} \cU_{[s]} \F ~, \qquad \bar{D}_\ad  \F = 0~, \\
	s > 0:&& \qquad ~~ \cU_{[s]} = \bar{D}^2 \bigg ( \z^{a(s) \b} (\partial_a)^s D_\b + \z^{a(s-1)} (\partial_a)^{s-1} D^2 \Big ) ~, \\
	\label{SGTF}
	s = 0:&& \qquad ~~	\cU_{[0]} = \bar{D}^2 \Big ( \z^{\a} D_\a \Big) ~,
	\eea
\end{subequations}
where $D_A = (\partial_a,D_\a,\bar{D}^\ad)$ are the covariant derivatives of Minkowski superspace and the gauge parameters $\z^{a_1 \dots a_s \b}$ and $\z^{a_1 \dots a_{s-1}}$ are symmetric in their vector indices and traceful.\footnote{We note that eq. \eqref{SGTF} corresponds to the transformation of $\F$ under the supergravity $\L$-supergroup, see e.g. \cite{Book}\ for more details.} 
This transformation law may be used to develop an alternative to the approach of \cite{KPR} to construct the 
interacting $\cN=1$ superconformal higher-spin theory. In such an alternative scheme, superconformal symmetry will not be manifest. 

In principle, one may extend the bosonic construction of \cite{BJM} to the $\cN=2$ superconformal case studied herein, however we favour the approach described above since it allows superconformal symmetry to remain manifest. It is worth pointing out that an $\cN=2$ extension of \eqref{4.7TL} was sketched in \cite{Buchbinder:2024pjm} within the framework of harmonic superspace.

\subsection{Induced action approach to nonlinear SCHS theory}

The results obtained in this work also open up the possibility to construct the nonlinear $\cN=2$ SCHS theory as an induced action in the spirit of the non-supersymmetric studies \cite{Tseytlin,Segal,BJM,Bonezzi,BeccariaT}. Recently, such an approach was utilised to compute the leading-order contribution to the nonlinear $\cN=1$ SCHS action in \cite{KLaP}. 

Specifically, once the analysis described in the previous subsection is completed, we can introduce an effective action $\G[\cH]$ associated with \eqref{4.1} according to
\begin{align}
	\label{EffectiveAction}
	\re^{\ri \G[\cH]} = \int \, \big [ \cD \U \big ] \, \big [ \cD \breve{\U} \big ] \, \re^{\ri \cS[\U,\breve{\U},\cH]} ~.
\end{align}
As $\cS[\U,\breve{\U},\cH]$ is bilinear in $\U^{(1)}$ and $\breve{\U}^{(1)}$, the latter may be integrated out in \eqref{EffectiveAction} to obtain $\G[\cH]$. The logarithmically divergent part of $\G[\cH]$ is local and gauge-invariant, thus, as sketched in \cite{KR21,KPR}, it is expected to coincide with the nonlinear SCHS theory. We anticipate that its leading contribution is given by the sum of linearised actions \eqref{SCHSaction}
\begin{align}
	S_\text{SCHS}[\cH] = \sum_{s=0}^{\infty}\bigg \{ \frac{(-1)^s}{4} \int \rd^4x \rd^4 \q \, \cE\, \mathbb{W}^{\a(2s+2)} \mathbb{W}_{\a(2s+2)} + \text{c.c.} \bigg \} + \cO(\cH^3)~,
\end{align}
where we have implicitly gauged away the auxiliary prepotentials; $\cH_\text{aux} = 0$.

\subsection{Harmonic-superspace construction} \label{section4.3}

To conclude, it is instructive to comment on the harmonic-superspace analogue of the projective-superspace construction we have presented, allowing us to directly compare our results with those of \cite{Buchbinder:2024pjm}. Our harmonic superspace conventions mostly agree with those of \cite{GIKOS}, with the notable exception that we denote by $\breve{q}^+$ the analyticity-preserving conjugate of $q^+$.

In a flat background, isotwistor supercurrents \eqref{IsoSC} may be translated to the following multiplets valued on $\mathbb{R}^{4|8} \times S^2$
\begin{align}
	\label{HSSSC}
	\mathfrak{J}^{\a(s) \ad(s)} &= \frac{\ri^s}{2} \sum_{k=0}^s (-1)^k \binom{s}{k}^2 \bigg \{ \frac{1}{k+1} (\partial^{\aa})^k D^{--} \breve{q}^{+} (\partial^\aa)^{s-k} q^{+} \non \\
	& \qquad\qquad - \frac{1}{s-k+1}(\partial^{\aa})^k \breve{q}^{+} (\partial^\aa)^{s-k} D^{--} q^{+} \non \\
	& \qquad\qquad - \frac{\ri}{2} \frac{(s-k)^2}{(k+1)(k+2)} (\partial^\aa)^k D^{\a-} \bar{D}^{\ad -} \breve{q}^{+} (\partial^\aa)^{s-k-1} q^{+} \non \\
	& \qquad\qquad - \frac{\ri}{2} \frac{s-k}{s-k+1} (\partial^\aa)^k \breve{q}^{+} (\partial^\aa)^{s-k-1} D^{\a-} \bar{D}^{\ad -} q^{+} \non \\
	& \qquad\qquad + \frac{\ri}{2} \frac{s-k}{k+1} (\partial^{\aa})^k D^{\a-} \breve{q}^{+} (\partial^\aa)^{s-k-1} \bar{D}^{\ad -}q^{+}  \non \\
	& \qquad\qquad - \frac{\ri}{2} \frac{s-k}{k+1} (\partial^{\aa})^k \bar{D}^{\ad -} \breve{q}^{+} (\partial^\aa)^{s-k-1} D^{\a-} q^{+} \bigg \}~.
\end{align}
Here $q^+$ is an analytic superfield
\begin{align}
	D_\a^+ q^+ = 0 ~, \qquad \bar{D}_\ad^+ q^+ = 0~,
\end{align}
describing, along with its conjugate $\breve{q}^+$, the off-shell hypermultiplet. On-shell, it obeys $D^{++} q^+ = 0$ and so the supercurrent \eqref{HSSSC} satisfies the conservation equation
\begin{align}
	\label{HarmonicConservationEq}
	D^{++} \mathfrak{J}^{\a(s) \ad(s)} = 0~,
\end{align}
implying that it is $u$-independent. Actually, they reduce to the (flat superspace limit of the) conformal supercurrents \eqref{HMSC}, which first appeared in \cite{KR21} for $s>0$ and \cite{KT} when $s=0$. Certain higher-spin supercurrents were also presented in \cite{Buchbinder:2024pjm}, however they do not coincide with \eqref{HSSSC}. In particular, they are not primary and are characterised by different conservation laws, see \cite{Buchbinder:2024pjm} for more details.

The higher-spin supercurrents given above are naturally dual to the primary gauge multiplets $\mathfrak{H}_{\a(s) \ad(s)}$, $s\geq0$, via the Noether coupling
\begin{align}
	\label{4.4}
	\mathfrak{S}_\text{NC} = \sum_{s=0}^{\infty} \int \rd u \,  \int \rd^{4|8}z \, \mathfrak{H}_{\a(s) \ad(s)} \mathfrak{J}^{\a(s) \ad(s)}~.
\end{align}
Conservation equation \eqref{HarmonicConservationEq} implies that these superfields are defined modulo
\begin{align}
	\label{4.5}
	\d_{\L} \mathfrak{H}_{\a(s) \ad(s)} = D^{++} \L_{\a(s) \ad(s)}^{--}~,
\end{align}
for unconstrained $\L_{\a(s) \ad(s)}^{--}$. Further, they enjoy the pre-gauge symmetry
\begin{subequations}
	\label{4.6}
	\bea
	s > 0:&& \qquad ~~ \d_\l \mathfrak{H}_{\a(s) \ad(s)} = \bar{D}_\ad^{+} \l^{-}_{\a(s) \ad(s-1)} + \text{s.c.} ~, \\
	s = 0:&& \qquad \quad\qquad~\,	\d_\l \mathfrak{H} = (\bar{D}^{+})^2 \l^{--} + \text{s.c.} \label{4.6b}
	\eea
\end{subequations}
Transformation law \eqref{4.6b} was first obtained in \cite{KT}. This symmetry originates similarly to its projective-superspace cousion \eqref{PreGT}. Specifically, by rewriting the Noether coupling \eqref{4.4} as an integral over analytic superspace
\begin{align}
	\label{NCHSS}
	\mathfrak{S}_\text{N.C.}= \sum_{s=0}^\infty \int {\rm d}\zeta^{(-4)}\, \breve{q}^{+} \mathfrak{H}^{++}_{[s]} q^{+}~, \qquad \mathfrak{H}^{++}_{[s]} \propto (D^{+})^4 \bigg( \mathfrak{H}^{\a(s) \ad(s)} (\partial_\aa)^s D^{--} + \dots \bigg)~,
\end{align}
one can check that $\mathfrak{H}^{++}_{[s]}$ is inert under \eqref{4.6}. The complete form of $\mathfrak{H}_{[s]}^{++}$ and proof of superconformal invariance of \eqref{NCHSS} will be discussed elsewhere.\footnote{An action of the form \eqref{NCHSS} was given in \cite{Buchbinder:2024pjm}, however their choice of $\mathfrak{H}^{++}_{[s]}$ differs from ours. In particular, given their choice of operator, action \eqref{NCHSS} is not manifestly superconformal. In our opinion, the analysis of superconformal invariance carried out in \cite{Buchbinder:2024pjm} is incomplete.}

The freedom \eqref{4.5} may be utilised to gauge away the infinite tail of $u$-dependent superfields in $\mathfrak{H}_{\a(s) \ad(s)}$ and impose the gauge
\begin{align}
	\label{4.7}
	D^{++} \mathfrak{H}_{\a(s) \ad(s)} = 0~.
\end{align}
In this gauge, these multiplets reduce to the SCHS gauge superfields \eqref{SCHSgt} proposed in \cite{HST}
\begin{align}
	\mathfrak{H}_{\a(s) \ad(s)} \xrightarrow{D^{++} \mathfrak{H}_{\a(s) \ad(s)} \,=\, 0}
	H_{\a(s) \ad(s)}~.
\end{align}
There is a family of combined gauge transformations of the form \eqref{4.5} and \eqref{4.6} preserving the gauge \eqref{4.7}. These are exactly the SCHS gauge transformations \cite{KR21,HST}
\begin{subequations}
	\bea
	s > 0:& \qquad ~~~~\d_{\l} H_{\a(s) \ad(s)} = \bar{D}_{\ad}^i \l_{\a(s) \ad(s-1) i} + \text{c.c.} ~, \\
	s = 0:& \qquad
	\d_{\l} H = \bar D^{ij} \l_{ij} + \text{c.c.} \label{4.20b}
	\eea
\end{subequations}
Transformation law \eqref{4.20b} was derived in \cite{KT} using the harmonic superspace approach to $\cN=2$ supergravity.
\\

\noindent
{\bf Acknowledgements:}\\
SK acknowledges the kind hospitality of Imperial College, London and the INFN, Sezione di Padova, during his visit in June--July 2024. 
The work of SK is supported in part by the Australian Research Council, project No. DP200101944.
The work of ER is supported by the Brian Dunlop Physics Fellowship.

\appendix

\section{$\cN=2$ superconformal higher-spin multiplets}
\label{AppendixB}

In this appendix we collect the results of \cite{KR21} which are essential to the studies undertaken in the main body.

\subsection{Superconformal higher-spin multiplets}
\label{AppendixB.1}

The conformal supercurrent multiplets $J^{\a(s) \ad(s)} $ reviewed in section \ref{IsotwistorCSCs}
 are naturally dual to the superconformal gauge multiplets $H_{\a(s) \ad(s)}$. This may be seen by considering the following Noether coupling
\bea
\label{A.5}
S_\text{NC} = \sum_{s=0}^{\infty}
\int \rd^{4|8}z \, E\, H_{\a(s) \ad(s) } J^{\a(s) \ad(s)} ~.
\eea
Specifically, it follows from the conservation laws \eqref{SuperC1} and \eqref{SuperC3} that the primary superfields $H_{\a(s)\ad(s)}$ are defined modulo the gauge transformations
\begin{subequations}
\label{SCHSgt}
\bea
\label{SCHSgt1}
s > 0:& \qquad ~~\d_{\l} H_{\a(s) \ad(s)} = \bar{\nabla}_{\ad}^i \l_{\a(s) \ad(s-1) i} + \text{c.c.} ~, \\
\label{SCHSgt2}
s = 0:& \qquad
\d_{\l} H = \bar \nabla^{ij} \l_{ij} + \text{c.c.}~,
\eea
\end{subequations}
where the gauge parameters $\l_{\a(s) \ad(s-1) i}$ and $\l_{ij}$ are complex unconstrained. In Minkowski superspace, gauge transformation law \eqref{SCHSgt1} was first proposed in \cite{HST}. Requiring the Noether coupling \eqref{A.5} to be locally superconformal uniquely fixes the dimension and $\sU(1)_R$ charge of $H_{\a(s) \ad(s)}$
\bea
\mathbb{D} H_{\a(s) \ad(s)} = - (s+2) H_{\a(s) \ad(s)} ~, \qquad {\mathbb Y} H_{\a(s) \ad(s)} = 0 ~.
\eea

By making use of the gauge transformations \eqref{SCHSgt}, a Wess-Zumino gauge can be constructed on $H_{\a(s)\ad(s)}$ to facilitate the study of its $\cN=1$ superfield or component content. Such studies and their results are available in \cite{KR21}, see also \cite{Thesis}.\footnote{For $s=0$, the $\cN=2 \rightarrow \cN=1$ reduction was carried out earlier \cite{BK2}.}

\subsection{Superconformal model for $H_{\a(s)\ad(s)}$}

From the prepotential $H_{\a(s)\ad(s)}$, we may construct the linearised higher-spin super-Weyl tensors\footnote{Actually, for $s=0$, \eqref{HSSW} is exactly the linearised super-Weyl tensor.}
\bea
	\label{HSSW}
	{\mathbb{W}}_{\a(2s+2)} = \bar \nabla^4 (\nabla_\a{}^{\bd})^s \nabla_{\a(2)} H_{\a(s) \bd(s)} ~, \quad \bar{\nabla}_\ad^i \mathbb{W}_{\a(2s+2)} = 0 ~, 
\eea
which are gauge-invariant in arbitrary conformally flat backgrounds
\begin{align}
	W_{\a \b} = 0 \quad \implies \quad \d_\l {\mathbb{W}}_{\a(2s+2)} = 0~.
\end{align}
Thus, in what follows we restrict our attention to supergeometries characterised by vanishing super-Weyl tensor, $W_{\a \b} = 0$.
In \eqref{HSSW} we have introduced 
the second-order operators 
\begin{align}
	\nabla_{\a\b} := \nabla_{(\a}^i \nabla_{\b) i} \ , \qquad
	\bar{\nabla}^{\ad\bd} := \bar\nabla^{(\ad}_i \bar\nabla^{\bd) i} \ .
\end{align}
Chiral field strength \eqref{HSSW} is characterised by the following superconformal properties:
\begin{align}
	K^B {\mathbb{W}}_{\a(2s+2)} = 0 ~, \qquad \mathbb{D} {\mathbb{W}}_{\a(2s+2)} = {\mathbb{W}}_{\a(2s+2)} ~, \qquad \mathbb{Y} {\mathbb{W}}_{\a(2s+2)} = -2 {\mathbb{W}}_{\a(2s+2)}~.
\end{align}

The above properties imply that the action
\bea
\label{SCHSaction}
S^{(s)}_\text{SCHS} = \frac{(-1)^s}{4} \int \rd^4x \rd^4 \q \, \cE\, \mathbb{W}^{\a(2s+2)} \mathbb{W}_{\a(2s+2)} + \text{c.c.} ~,
\eea
is locally superconformal and gauge-invariant. It is normalised in accordance with the identity
\bea
\ri \int \rd^4x \rd^4 \q \, \cE\, \mathbb{W}^{\a(2s+2)} \mathbb{W}_{\a(2s+2)} + \text{c.c.} = 0~,
\eea 
which holds up to a total derivative. The free theory \eqref{SCHSaction} is known to possess $\sU(1)$ duality invariance, see \cite{KR21-2,KR23-2}. By making use of the formalism for $\sU(1)$ duality rotations, nonlinear extensions of \eqref{SCHSaction} were constructed in \cite{KR21-2}.

\begin{footnotesize}

\end{footnotesize}

\end{document}